\documentclass[12pt]{iopart}
\usepackage{graphicx}
\usepackage{xcolor}
\usepackage{stackrel}
\begin{document}
\title[Phase Sensitive Amplification Enabled by Coherent Population Trapping]{Phase Sensitive Amplification Enabled by Coherent Population Trapping}

\author{P. Neveu, C. Banerjee, 
J. Lugani\footnote{Present address:
Clarendon Laboratory, University of Oxford, Parks Road, Oxford OX1 3PU, UK.}, F. Bretenaker,  E. Brion and F. Goldfarb}

\address{Laboratoire Aim\'e Cotton, Universit\'e Paris-Sud, ENS Paris-Saclay,
CNRS, Universit\'e Paris-Saclay, 91405 Orsay, France}

\begin{abstract}
We isolate a novel four-wave mixing process, enabled by Coherent Population Trapping (CPT), leading to efficient phase sensitive amplification. This process is permitted by the exploitation of two transitions starting from the same twofold degenerate ground state. One of the transitions is used for CPT, defining bright and dark states from which ultra intense four-wave mixing is obtained via the other transition. This leads to the measurement of a strong phase sensitive gain even for low optical densities and out-of-resonance excitation. The enhancement of four-wave mixing is interpreted in the framework of the dark-state polariton formalism.
\end{abstract}

\submitto{\NJP}
\maketitle

\section{Introduction}
Optical parametric amplification processes have been widely studied for their unique noise properties and their many possible applications in metrology \cite{McKenzie02}, imaging \cite{Sokolov04} and telecommunications \cite{Lundstrom11}. They have thus been implemented in different media such as nonlinear crystals and waveguides \cite{Levenson93} through
three-wave mixing ($\chi^{(2)}$ process) or fibers \cite{Tong12} through four-wave mixing ($\chi^{(3)}$ process) (FWM): one or two strong driving pump field(s) play the role of a reservoir of photons for a signal and an idler fields. Depending on the relative phase between these fields, photons can be transferred from the pump(s) to the signal and idler fields or conversely. Such noiseless phase sensitive amplification (PSA) processes allow for the amplification of a coherent state of light  into another minimum uncertainty state, keeping the product of the field quadratures variances constant \cite{Walls83}. This is associated with the generation of squeezed states of light, which are of interest for quantum optics, atomic memories, entanglement swapping, and quantum information processing protocols \cite{Agarwal13}. Very large quantum noise reductions up to 15 dB have been achieved using crystals \cite{Vahlbruch16}, but down-converted photons are spectrally mismatched with atomic systems used for quantum memories and PSA achieved directly through FWM in atomic systems like alkali vapors is a subject of active interest \cite{Corzo12,Jing15,Jing16}.

In atomic systems, FWM efficiency can be boosted up using coherent population trapping (CPT) \cite{Harris90}. This two-photon process arises in a $\Lambda$-system and suppresses the absorption of a light field even at optical resonance by optically pumping the population into a dark state, which is a coherent superposition of two states. Consequently, this linear absorption suppression makes multiphoton processes such as $\chi^{(3)}$ processes predominant. Theoretical proposals based on CPT enhancement of FWM were put forward in double-$\Lambda$ systems \cite{Dantan03} and experimental implementations were also reported in rubidium \cite{Wang00,Lu98}, sodium \cite{Hemmer95}, and cesium \cite{Wang17}. 

\begin{figure}
\begin{centering}
\includegraphics[width=0.85\columnwidth]{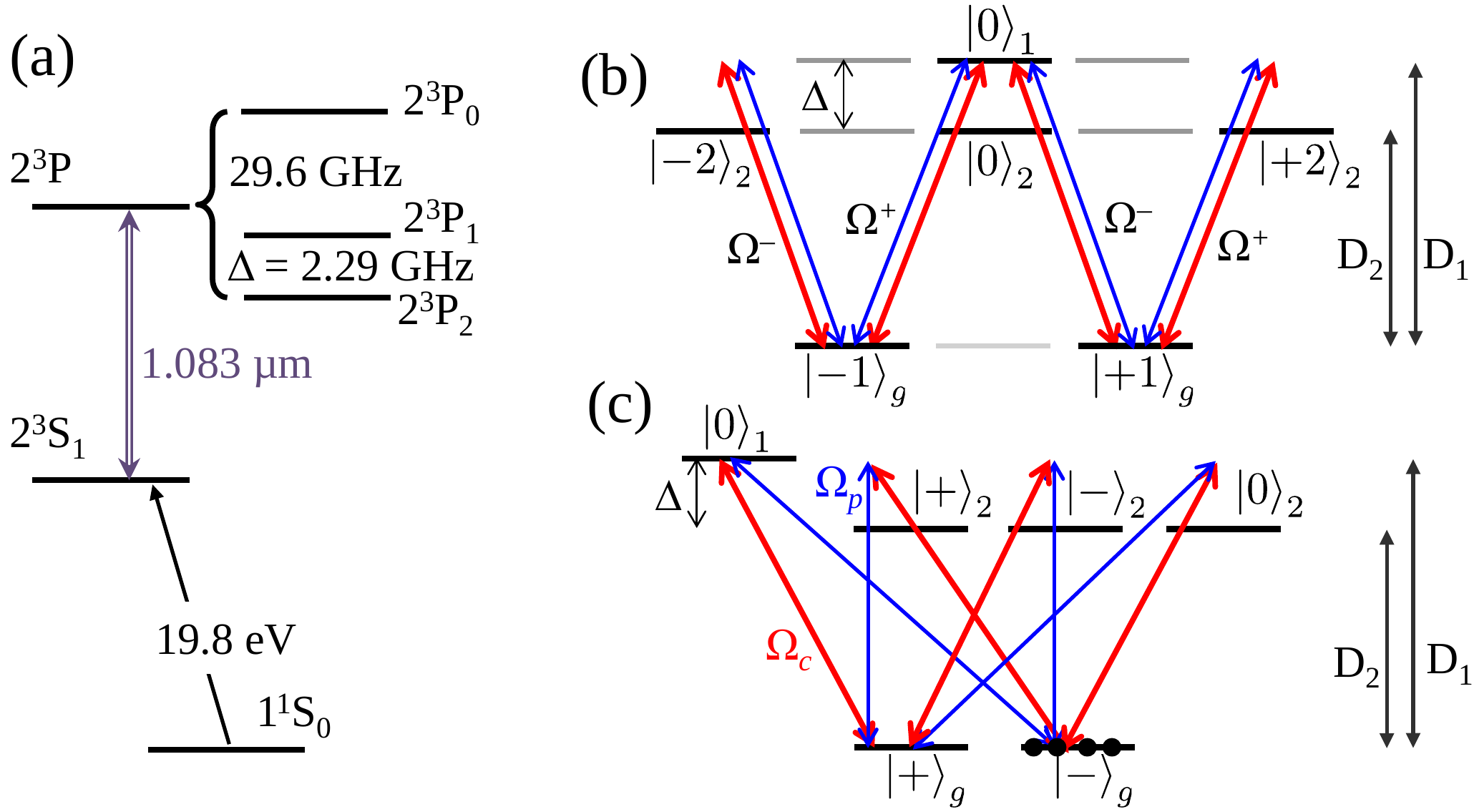}
\par\end{centering}
\caption{(a) Level scheme in helium 4. The $D_{1}$ transition ($2^{3}S_{1}\leftrightarrow2^{3}P_{1}$) is resonantly excited, while the $D_{2}$ transition ($2^{3}S_{1}\leftrightarrow2^{3}P_{2}$)
is far detuned by $\Delta$ (the $^{3}P_{0}$ level is far away and can be overlooked): excitation schemes are shown in the atomic basis $\mathcal{B}_{\mathrm{at}}$ (b) and in the dark and bright
states basis $\mathcal{B}_{\mathrm{s}}$ defined by the coupling field  (c). The $^{3}S_{1}$, $^{3}P_{1}$ and $^{3}P_{2}$ states are labeled by the indices $g$, $1$ and $2$ respectively. Due to
selections rules and optical pumping, grey-shadowed Zeeman levels can be neglected and the relevant ones are labeled using their $m$ numbers. CPT occurs through the $D_{1}$ transition, pumping the population into the $\left|-\right\rangle _{g}$ state, thus inducing full transparency and enabling for efficient multiphoton processes.\label{System}}
\end{figure}
However, because of their hyperfine structure, alkali atoms do not offer convenient closed $\Lambda-$systems. In this paper, following the experimental results of \cite{Lugani16}, we exploit the simple level structure of metastable helium 4: because of selection rules, the $D_{1}$ transition constitutes a well defined closed lambda system, allowing for a strong CPT effect to occur. Two other transitions share the same ground states, which can then be fully exploited to have multiphoton nonlinear processes explicitly addressing the dark and bright states. Therefore, we expect this atom to exhibit a strong nonlinear third-order susceptibility while being free from absorption. We moreover develop an analytic treatment to extract the properties of the amplification process, and show that it has the properties of a perfect squeezer \cite{Ferrini14}.

\section{Experiment and modelling}

The relevant level structure of helium is shown in Fig.\,\ref{System}(a): the long lived $2^{3}S_{1}$ metastable state is populated using a radio-frequency discharge, and optically coupled to the $2^{3}P$ fine states at wavelengths close to  $1.083\,\mu$m. The upper level $^{3}P_{0}$  is separated from the $^{3}P_{1}$ states by more than 20 times the Doppler broadening $W\simeq2\pi\times0.9\,\mathrm{GHz}$ and can thus be overlooked. The decay rate $\Gamma$ of the optical coherence at room temperature in the $1\,\mathrm{Torr}$ helium cell is about $2\pi\times23\,\mathrm{MHz}$, but following \cite{Figueroa06,Goldfarb08}, its value is replaced by the Doppler width $W$ in the simulations performed below. 

\begin{figure}
\begin{centering}
\includegraphics[width=\columnwidth]{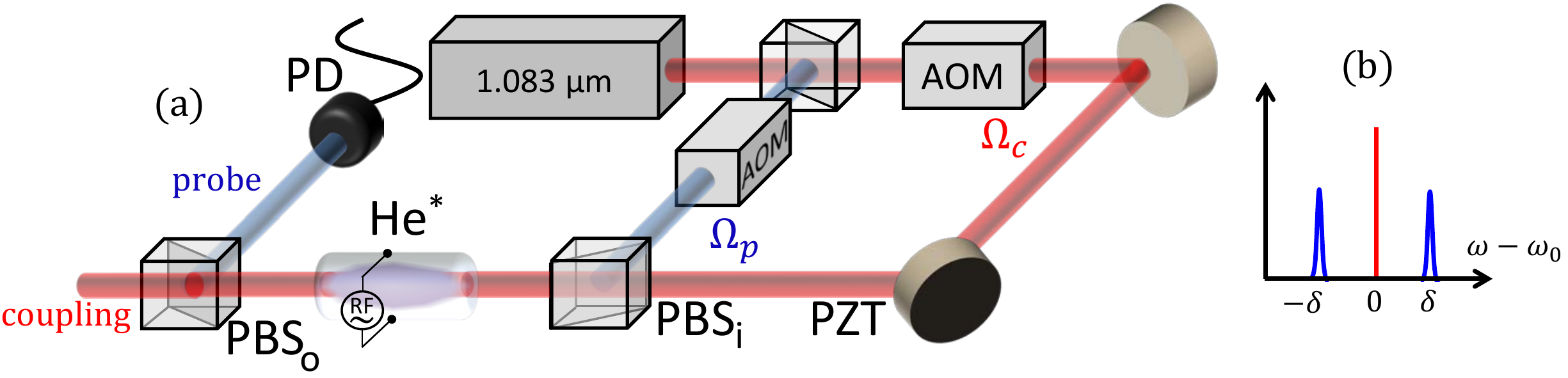}
\includegraphics[width=\columnwidth]{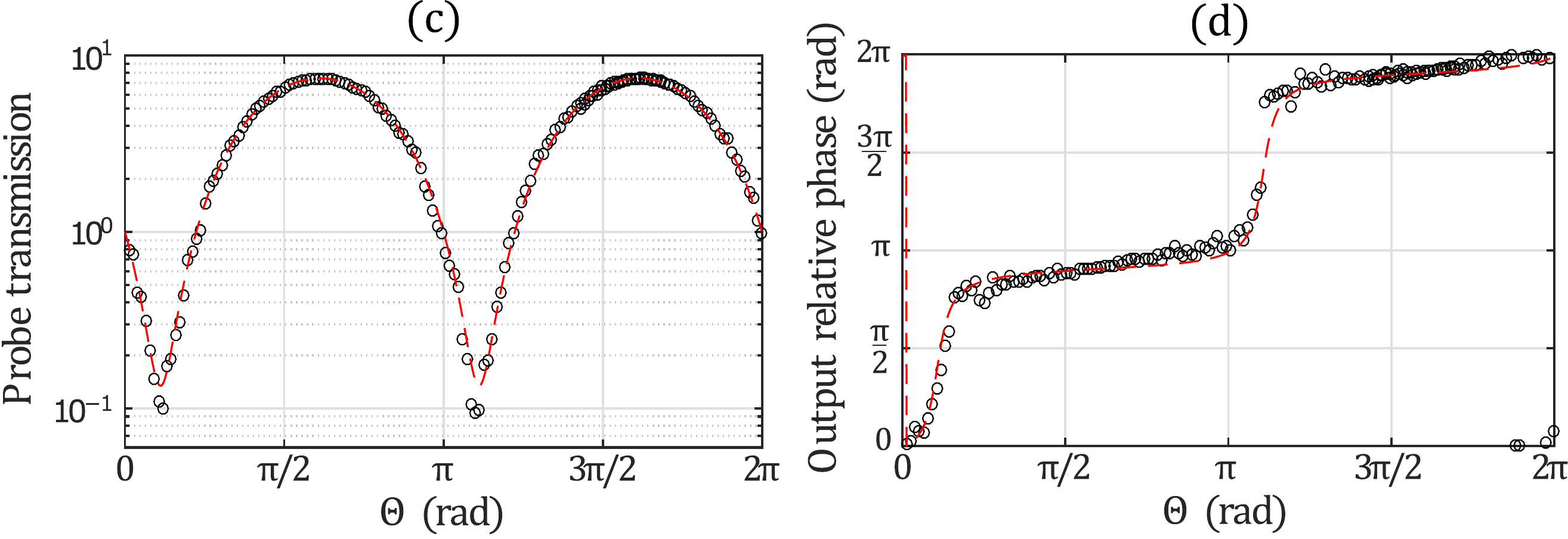}
\par\end{centering}
\caption{(a) Experimental setup. A laser diode at $1.083\,\mu$m is used to generate orthogonally and linearly polarized probe and coupling fields $\Omega_{p}$ and $\Omega_{c}$, the latter one being amplified by a tapered amplifier. Acousto-optic modulators (AOM) enable to generate arbitrary spectra for the fields, such as the typical degenerate pump PSA scheme represented in (b). Two electrodes  placed apart the cell generate a breakdown voltage at 27\,MHz radiofrequency (RF): collisions with the electrons of the plasma then pump the atoms to the metastable state $2^3S_1$. The input relative phase $\Theta$ between signal ($\omega_{0}+\delta$), idler ($\omega_{0}-\delta$), and coupling ($\omega_{0}$) fields is tuned using a piezoelectric transducer (PZT). The sideband signal and idler fields transmission through the $6\,$cm long cell is measured by a photodiode (PD). Input (output) relative phase between probe and coupling fields is measured via a small leakage of the fields at the cell input beamsplitter $\mathrm{PBS_i}$ (the coupling field at the cell output beamsplitter $\mathrm{PBS_o}$). (c) When the relative phase is scanned with $\delta=2\pi\times2\:\mathrm{kHz}$ , one observes a PSA of the sidebands fields with a maximum gain equal to 7.4 ($8.7\,\mathrm{dB}$) and (d) the associated stabilization of the probe field relative phase with the coupling field.  Gains up to $9.3\,\mathrm{dB}$ have been measured. Open circles: measurements. Dashed lines: fit with $\mu=1.18$ using Eq.\,(\ref{eq:Matrix}).} \label{Exp-1}\end{figure}

The experimental set-up is schemed in Fig.\,\ref{Exp-1}(a). The $2^{3}S_{1}\leftrightarrow2^{3}P_{1}$ ($D_{1}$) helium transition is resonantly excited by a strong $200\,\mathrm{W.cm}^{-2}$ coupling field and a weak $0.50\,\mathrm{W.cm}^{-2}$ probe field, of respective Rabi frequencies $\Omega_{c}$ and $\Omega_{p}$, with $\left|\Omega_{p}\right|\ll\left|\Omega_{c}\right|$. 
The coupling and probe fields are orthogonally and linearly polarized so that the Rabi frequencies involved in the circularly polarised light basis $\sigma^{\pm}$ are:
\begin{equation}
\Omega^{\pm}=\frac{1}{\sqrt{2}}\left(\Omega_{c}\pm\mathrm{i}\Omega_{p}\right)\ .
\end{equation}
In PSA configuration, the probe field contains two frequencies, separated by $\pm\delta$ from the coupling field frequency (see Fig.\,\ref{Exp-1}(b)), and called signal and idler. A typical probe transmission experimental measurement in PSA configuration is reproduced in Fig.\,\ref{Exp-1}(c): a maximum gain equal to 9.3 dB is observed for the probe field.

As the transition $2^{3}S_{1}(m=0)\leftrightarrow2^{3}P_{1}(m=0)$ is forbidden, the resonant interaction with the fields leads to a $\Lambda$-type level scheme for the $D_{1}$ transition (see Fig.\,\ref{System}(b)), where the $^{3}S_{1}$, $^{3}P_{1}$ and $^{3}P_{2}$ states are labeled by the indices $g$, $1$ and $2$ respectively. Because the linearly polarised excitation of this $\Lambda$-system leads to a strong CPT effect, which suppresses the $D_1$ transition linear absorption, one needs to take into account the nearest transition to find the most efficient multiphoton processes \cite{Harris90}. The Zeeman sublevels involved in the different processes are labeled by their $m$ magnetic quantum numbers: $\left|\pm1\right\rangle _{g}=\left|2^{3}S_{1},m=\pm1\right\rangle $ and $\left|0\right\rangle _{1}=\left|2^{3}P_{1},m=0\right\rangle $ are the relevant ground and excited states of the $D_{1}$ transition, while $\left|\pm2\right\rangle _{2}=\left|2^{3}P_{2},m=\pm2\right\rangle $ and $\left|0\right\rangle _{2}=\left|2^{3}P_{2},m=0\right\rangle $ are coupled to the ground states $\left|\pm1\right\rangle _{g}$ by the $D_{2}$ transition. The population of the $\left|2^{3}S_{1},m=0\right\rangle $ state through the far-detuned $2^{3}S_{1}\leftrightarrow2^{3}P_{2}$ ($D_{2}$) transition can be neglected ($\Delta/2\pi=2.29\,\mathrm{GHz}$), and therefore the $\left|2^{3}P_{2},m=\pm1\right\rangle $ states are also not relevant. In the rotating wave approximation, the interaction Hamiltonian $H$ in the atomic basis $\mathcal{B}_{\mathrm{at}}=\left\{ \left|0\right\rangle _{1},\left|-1\right\rangle _{g},\left|+1\right\rangle _{g},\left|0\right\rangle _{2},\left|-2\right\rangle _{2},\left|+2\right\rangle _{2}\right\} $, is given by:
\begin{equation}
\left(\begin{array}{cccccc}
0 & \Omega^{+} & \Omega^{-} & 0 & 0 & 0\\
\Omega^{+*} & 0 & 0 & -\frac{\Omega^{+*}}{\sqrt{3}} & \sqrt{2}\Omega^{-*} & 0\\
\Omega^{-*} & 0 & 0 & \frac{\Omega^{-*}}{\sqrt{3}} & 0 & -\sqrt{2}\Omega^{+*}\\
0 & -\frac{\Omega^{+}}{\sqrt{3}} & \frac{\Omega^{-}}{\sqrt{3}} & \Delta & 0 & 0\\
0 & \sqrt{2}\Omega^{-} & 0 & 0 & \Delta & 0\\
0 & 0 & -\sqrt{2}\Omega^{+} & 0 & 0 & \Delta
\end{array}\right)\ ,
\end{equation}
where the numerical factors originate from the Clebsch-Gordan coefficients.

One can then derive the evolution of the density matrix $\rho$ via optical Bloch equations:
\begin{equation}
\mathrm{i}\hbar\partial_{t}\rho=\left[H,\rho\right]+\mathcal{L}\left(\rho\right)\ ,
\label{VN}
\end{equation}
where $\mathcal{L}$ stands for the non Hermitian dynamics caused by spontaneous emission and extra optical coherence decay, of rates  $\Gamma_{0}$ and $\Gamma$ respectively. The fields along $z$ then propagate according to Maxwell's equations in the slowly varying envelope approximation:
\begin{equation}
\left(c\partial_{z}+\partial_{t}\right)\Omega^{\pm}=\mathrm{i}c\eta\left(\pm\frac{1}{\sqrt{3}}\rho_{0_{2}\mp1_{g}}\pm\sqrt{2}\rho_{\pm2_{2}\pm1_{g}}-\rho_{0_{1}\mp1_{g}}\right)\ , \label{eq04}
\end{equation}
where $\eta$ is the atom-field coupling coefficient, $\rho_{ij}=\mathrm{Tr}\left[\rho\left|i\right\rangle \left\langle j\right|\right]$,
and numerical factors are given by Clebsch-Gordan coefficients. 
\section{Results}
\label{Section3}
\begin{figure}
\begin{centering}
\includegraphics[width=1\columnwidth]{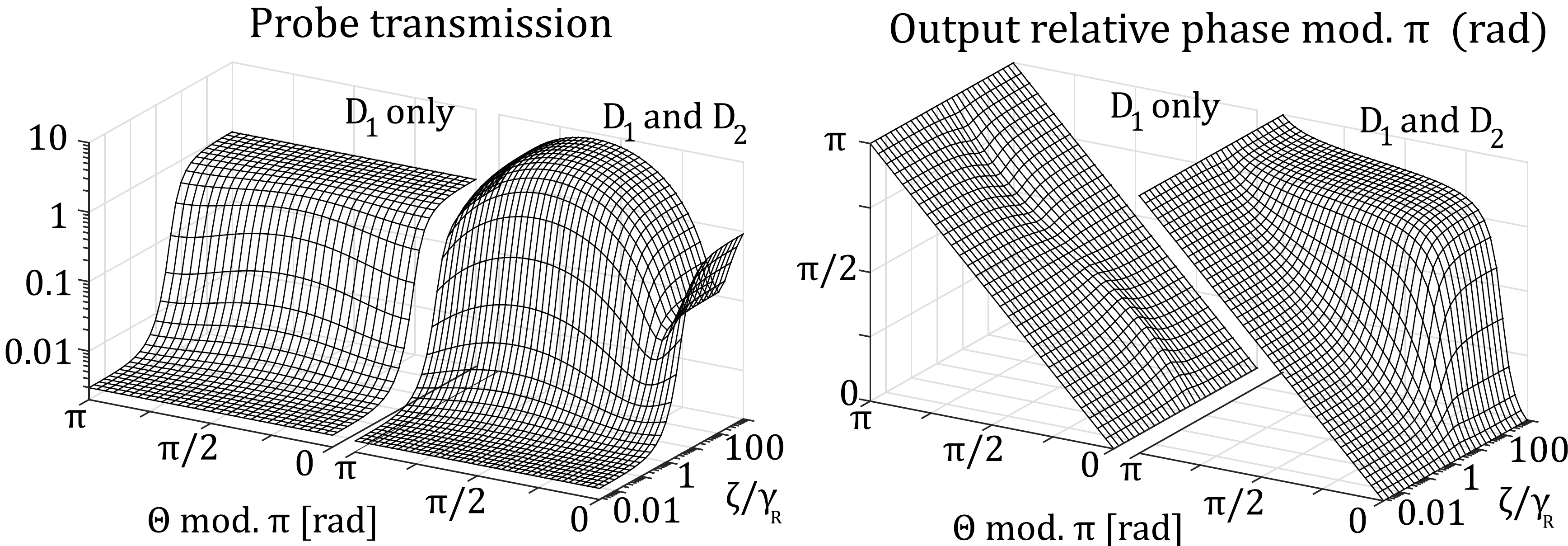}
\par\end{centering}
\caption{(a) Simulated transmission of the probe field based on the Maxwell-Bloch formalism when the fields are degenerate ($\delta=0$), not taking (right) and taking (left) into account the $D_{2}$ transition. All the parameters such as optical depth and decay rates corresponds to the ones measured experimentally. The probe transmission is plotted as a function of its initial relative phase $\Theta$ with the coupling field and of the coupling field strength normalized to the Raman coherence decay rate $\zeta/\gamma_{R}$. When CPT is efficient ($\zeta/\gamma_{R}\gg1$), the $D_{1}$ transition becomes transparent and PSA occurs via the $D_{2}$ levels. The value of the optical depth of the medium used for this plot is extracted from experimental measurements. (b) Evolution of the output relative phase between the probe and the coupling field in the same conditions. When CPT is efficient ($\zeta/\gamma_{R}\gg1$), PSA induces a stabilization of the output relative phase to the value $\Theta_{\mathrm{MAX}}$.\label{Simu}}
\end{figure}

Fig.\,\ref{Simu}(a) shows the result of the numerical simulations of the probe field intensity transmission coefficient in the degenerate case ($\delta=0$), as a function of its relative phase $\Theta$ with respect to the coupling field and the  power-broadening factor $\zeta=\Omega_{c}^{2}/\Gamma$ of the coupling field normalized to the Raman coherence decay rate $\gamma_R$ between the levels $\left|\pm1\right\rangle _{g}$. For clarity, we compare a numerical simulation where the $D_{1}$ transition is considered alone (right panel) with another where the $D_{1}$ and the $D_{2}$ transitions are both considered (left panel). In both cases, CPT occurs when the coupling field strength overcomes the Raman coherence decay rate $\gamma_{R}$. Below this threshold (i.e., for $\zeta/\gamma_{R}\ll1$), the resonant absorption by the $D_{1}$ transition forbids any multiphoton process. Above this threshold, (i.e., for $\zeta/\gamma_{R}\gg1$), CPT becomes efficient: PSA then occurs with a gain as large as 8.5 for an optical depth of 4.5 only when the $D_2$ line is taken into account while, whatever the phase is, the probe transmission remains 1 for large values of $\zeta$ when the $D_2$ transition is overlooked. Moreover, Fig.\,\ref{Simu}(b) shows the evolution of the output relative phase in the same conditions. In the regime where CPT does not exist ($\zeta/\gamma_{R}\ll1$), the phase is unchanged through propagation. However, when CPT exists ($\zeta/\gamma_{R}\gg1$) the output relative phase is stabilized to the specific value $\Theta_{\mathrm{MAX}}$, as experimentally observed in Fig.\,\ref{Exp-1}(d). These simulations are computed with decay rates  parameters which correspond to experimental measurements.

Let us now focus on the degenerate case $\delta=0$ and assume $\zeta\gg\gamma_{R}$. In the steady state regime, restricting Eq. (\ref{eq04}) for $\Omega_{c}$ and $\Omega_{p}$ to the leading order terms in $\Omega_{p}/\Omega_{c}$ leads to:

\begin{equation}
\partial_{z}\Omega_{c}=  \frac{4\mathrm{i}\eta}{3\Delta}\Omega_{c}+{\cal O}\left(\frac{\Gamma,\zeta}{\Delta}\right),
\label{eq:PropagStationnaireC}
\end{equation}
\begin{equation}
\partial_{z}\Omega_{p}=  \frac{4\mathrm{i}\eta}{3\Delta}\frac{2\Omega_{p}\Omega_{c}^{*}-\Omega_{c}\Omega_{p}^{*}}{\Omega_{c}^{*}}+{\cal O}\left(\frac{\Gamma,\zeta}{\Delta}\right),\label{eq:PropagStationnaireS}
\end{equation}
where the terms ${\cal O}\left(\Gamma,\zeta/\Delta\right)$ contain multiphoton processes involving several times the $D_{2}$ far-detuned transition.

Equation (\ref{eq:PropagStationnaireC}) yields $\Omega_{c}\left(L\right)=\Omega_{c}\left(0\right)\exp\left(\mathrm{i}\mu\right)$ where $\mu=\frac{4\eta}{3\Delta}L$ and $L$ is the length of the atomic medium: the coupling field experiences a phase shift along its propagation because of the far-detuned $D_{2}$ transition. Here, the coupling field depletion by the probe field is neglected due to the first order approximation in $\Omega_{p}/\Omega_{c}$. This expression can then be used to solve Eq. (\ref{eq:PropagStationnaireS}), leading to:
\begin{equation}
\left(\begin{array}{c}
\Omega_{p}\left(L\right)\\
\Omega_{p}^{*}\left(L\right)
\end{array}\right)=\left(\begin{array}{cc}
\left(1+\mathrm{i}\mu\right)\mathrm{e}^{\mathrm{i}\mu} & \mathrm{i}\mu\, \mathrm{e}^{\mathrm{i}\mu}\\
-\mathrm{i}\mu\, \mathrm{e}^{-\mathrm{i}\mu} & \left(1-\mathrm{i}\mu\right)\mathrm{e}^{-\mathrm{i}\mu}
\end{array}\right)\left(\begin{array}{c}
\Omega_{p}\left(0\right)\\
\Omega_{p}^{*}\left(0\right)
\end{array}\right)\ ,\label{eq:Matrix}
\end{equation}
where one recognizes a typical PSA transfer matrix belonging to the symplectic group \cite{Ferrini14}. It provides a gain $G\left(\Theta\right)\equiv\left|\Omega_{p}\left(L,\Theta\right)/\Omega_{p}\left(0\right)\right|^{2}$
of maximum and minimum values $G_{\mathrm{MAX}}$ and  $G_{\mathrm{MIN}}$:
\begin{equation}
G_{\mathrm{MAX}}=1+2\mu\left(\mu+\sqrt{1+\mu^{2}}\right)=1/G_{\mathrm{MIN}}\ ,
\end{equation}
where the values $\Theta_{\mathrm{MAX}}$ and $\Theta_{\mathrm{MIN}}$ of the initial relative phase $\Theta$ between the fields are given by:
\begin{equation}
\Theta_{\mathrm{MAX}}=\frac{1}{2}\arctan\left(\frac{1}{\mu}\right)=\frac{\pi}{2}+\Theta_{\mathrm{MIN}}\quad\left[\pi\right] \ .
\end{equation}
At the quantum level, the properties of such a non-unitary transfer
matrix ensure that no extra noise is added during the amplification
process, leading to squeezed state generation.

\begin{figure}
\begin{centering}
\includegraphics[width=\columnwidth]{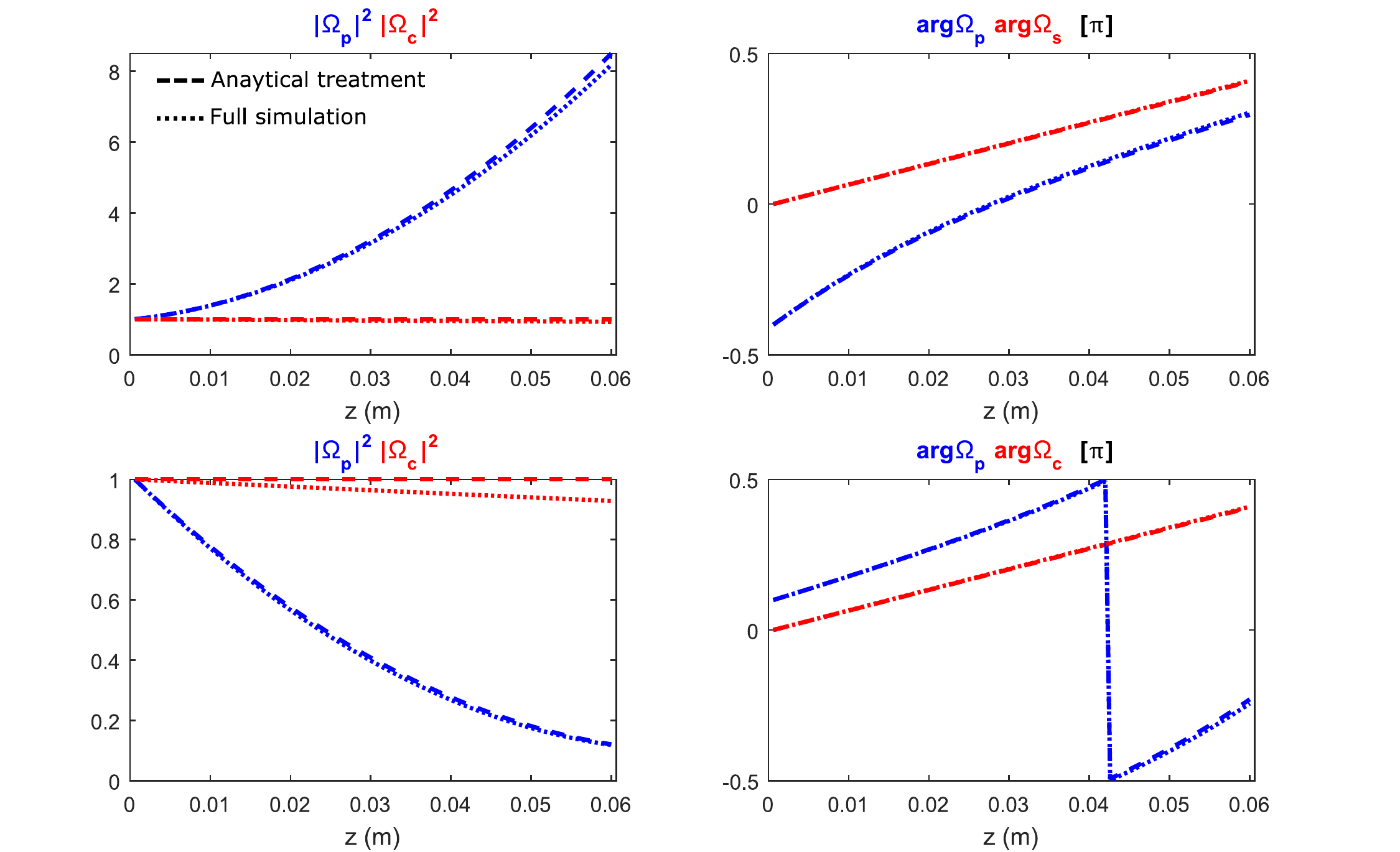}
\par\end{centering}
\caption{Plot of the intensity (left) of the probe and coupling pump field as well
as their phase (right) for the phase matching conditions $\Theta_{\mathrm{MAX}}$
(top) and $\Theta_{\mathrm{MIN}}$ (bottom). Intensities are normalized to their initial input value.}
\label{Figure5}
\end{figure}

The validity of our model to describe the experimental results is shown on Figs.\,\ref{Exp-1}c and \ref{Exp-1}d. The broken lines fit the experimental data with gain and output phase transfer functions that are derived from Eq.\,\ref{eq:Matrix}. Finally, we investigate the validity of our approximation framework by comparing the analytical results with the full simulation of the Maxwell-Bloch formalism. Fig.\,\ref{Figure5} displays the intensities and phases of the probe and coupling fields during propagation, for an initial relative phase $\Theta_\mathrm{MAX}$ ($\Theta_\mathrm{MIN}$) corresponding to a maximal gain $G _\mathrm{MAX}$ (minimal gain $G _\mathrm{MIN}$). Despite an excellent agreement, a small discrepancy is noticeable in particular on the coupling field intensity. Indeed, some residual $D_2$ absorption occurs, which is not taken into account in the analytical treatment.

Contrary to usual far off-resonance FWM schemes, the gain scales here as $1/\Delta$. Moreover, contrary to the usual  PSA behaviour, the maximum reachable gain $G_{\mathrm{MAX}}$ does not explicitly depend on the coupling field intensity but only on $\mu$, which is proportional to the optical depth of the medium. Indeed, as shown in Fig.\,\ref{Simu}, this process is possible only when the atoms are pumped into the dark state, which occurs when $\zeta\gg\gamma_{R}$ and does not result from any strong saturation effect. 
\section{Discussion}
To understand the underlying mechanism, it is interesting to switch to the CPT dark ($\left|-\right\rangle _{g}$) and bright ($\left|+\right\rangle _{g}$) state basis defined by the coupling field, which assigns different transitions to the coupling and probe fields \cite{Laupretre09}:
\begin{equation}
\left|\pm\right\rangle _{g}=\frac{\left|+1\right\rangle _{g}\pm\left|-1\right\rangle _{g}}{\sqrt{2}}\ .
\end{equation}
When the Zeeman sublevels are degenerate, the coupling (probe) field couples the  $\left|+\right\rangle _{g}$ ($\left|-\right\rangle _{g}$) state to the $\left|0\right\rangle _{e}$ state. Optical pumping in the $\left|-\right\rangle _{g}$ state suppresses the linear absorption by the $D_{1}$ transition, which then constitutes a highly efficient multiphoton channel \cite{Harris90} and allows for efficient nonlinearities with a far-detuned transition, such as the $D_{2}$ one. 

Following the same idea, the $D_{2}$ transitions shared by the coupling and probe fields can be decoupled if we use superpositions of the $D_{2}$ line upper levels:
\begin{equation}
\left|\pm\right\rangle _{2}=\frac{\left|+2\right\rangle _{2}\pm\left|-2\right\rangle _{2}}{\sqrt{2}}\ .
\end{equation}
Fig.\,\ref{System}(c) shows the relevant transitions in the basis $\mathcal{B}_{\mathrm{s}}=\left\{ \left|0\right\rangle _{1},\left|-\right\rangle _{g},\left|+\right\rangle _{g},\left|0\right\rangle _{2},\right.$ $\left.\left|-\right\rangle _{2},\left|+\right\rangle _{2}\right\} $. As soon as the population is trapped into the dark state $\left|-\right\rangle _{g}$ , we expect that only the processes involving the $D_2$ transition once and the $D_1$ transition once would play a significant role, namely:
\begin{equation}
\begin{array}{ll}
\mbox{FWM via \ensuremath{\left|\pm2\right\rangle _{2}}}: & \left|-\right\rangle _{g}\stackrel{\Omega_{c}}{\rightarrow}{}\left|+\right\rangle _{2}\stackrel{\Omega_{p}^{*}}{\rightarrow}\left|+\right\rangle _{g}\stackrel{\Omega_{c}}{\rightarrow}\left|0\right\rangle _{1}\stackrel{\Omega_{p}^{*}}{\rightarrow}\left|-\right\rangle _{g},\\
\mbox{FWM via \,\,\ensuremath{\left|0\right\rangle_{2} }:} & \left|-\right\rangle _{g}\stackrel{\Omega_{c}}{\rightarrow}\,\left|0\right\rangle _{2}\,\stackrel{\Omega_{p}^{*}}{\rightarrow}\left|+\right\rangle _{g}\stackrel{\Omega_{c}}{\rightarrow}\left|0\right\rangle _{1}\stackrel{\Omega_{p}^{*}}{\rightarrow}\left|-\right\rangle _{g}.
%  \mbox{FWM via \ensuremath{\left|\pm2\right\rangle _{2}}}: & \left|-\right\rangle _{g}\rightarrow\left|+\right\rangle _{2}\rightarrow\left|+\right\rangle _{g}\rightarrow\left|0\right\rangle _{1}\rightarrow\left|-\right\rangle _{g},\\
%  \mbox{FWM via \ensuremath{\left|0\right\rangle_{2} }:} & \left|-\right\rangle _{g}\rightarrow\left|0\right\rangle _{2}\rightarrow\left|+\right\rangle _{g}\rightarrow\left|0\right\rangle _{1}\rightarrow\left|-\right\rangle _{g}.
\end{array}
\end{equation}
FWM processes involving the $D_{2}$ transition twice have been neglected: they are much less efficient than the two processes cited above, which exploit the full transparency of the resonant $D_{1}$ transition. These two processes, enabled by CPT, correspond to the transfer matrix of Eq. (\ref{eq:Matrix}) and lead to the high PSA experimentally observed in \cite{Lugani16} and in Fig.\,\ref{System}(c).

As predicted by Fig.\,\ref{Simu}, the fact that the dominant FWM processes start from the dark state $\left|-\right\rangle _{g}$ implies that the $D_{1}$ line absorption destroys any multiphoton process when the coupling field is too weak: due to insufficient CPT, the population is incoherently shared between the ground states.

Let us now consider the more general case of a probe field with a finite initial spectrum $\Omega_{p}\left(z=0,\nu\right)$. The frequency $\nu=\omega-\omega_0$ is defined with respect to the monochromatic coupling field. Assuming that the probe spectrum fits within the $D_{1}$ transition linewidth, i.e. $\nu\ll\Gamma\ll\Delta$, the propagation equation for $\Omega_{p}\left(z,\nu\right)$
is:
\begin{equation}
\left(\partial_{z}+\mathrm{i}\frac{\nu}{\beta}\right)\Omega_{p}=  \frac{4\mathrm{i}\eta}{3\Delta}\frac{2\Omega_{p}\Omega_{c}^{*}-\Omega_{p}^{*}\Omega_{c}}{\Omega_{c}^{*}}f^{2}\left(\nu/\zeta\right)+{\cal O}\left(\frac{\Omega_{c,p},\Gamma,\zeta}{\Delta}\right)\ ,\label{eq12}
\end{equation}
where
\begin{equation}
\beta=\frac{c}{1+\frac{2\eta c}{\left|\Omega_{c}\right|^{2}}f\left(\nu/\zeta\right)}\,\,\,\,\textrm{ and }\,\,\,\,f\left(x\right)=\frac{1}{1-\mathrm{i}x}\ .
\end{equation}

Eqs. (\ref{eq12}) and (\ref{eq:PropagStationnaireS}) have the same right-hand side, provided the signal spectrum fits within the saturation-broadened CPT linewidth, i.e. $\nu\ll\zeta$.  Furthermore, $f\left(\nu/\zeta\right)\simeq1$ in this regime, and one can extract the probe field group velocity $v_{g}=c/(1+2\eta c/\left|\Omega_{c}\right|^{2})$. One recognizes the usual slow-light behaviour due to the strong dispersion created by the CPT narrow transparency window \cite{Fleischhauer05}.

Up to a redefinition of its phase, the probe field and its complex conjugate play a symmetric role in the propagation equation (see \ref{Appendix}), indicating that a signal detuned from the coupling field requires an idler input with a symmetric spectrum with respect to the coupling field frequency. For example, $\Omega_{p}$ can be the superposition of a signal and an idler fields peaked at $\pm\delta$ as represented in Fig.\,\ref{Exp-1}(b):
\begin{equation}
\Omega_{p}\left(0,\nu\right)=\mathcal{E}_{p}\left(0\right)\left[\delta_D\left(\nu-\delta\right)+\delta_D\left(\nu+\delta\right)\right]\ ,
\end{equation}
where $\delta_D$ is the Dirac distribution. The FWM process involves the stimulated emission of one idler and one signal photons, and the transfer matrix of the total signal and idler fields is thus symplectic like in the degenerate configuration.
 
When the probe spectrum fits within the CPT bandwidth, the dark state polariton (DSP) ${\cal P}$ can be introduced \cite{Fleischhauer00}
\begin{equation}
{\cal P}=\cos(\alpha)e^{-\frac{4\mathrm i \eta z}{3\Delta}}\Omega_{p}-\sqrt{2\eta c}\,\mathrm{i}\sin(\alpha)\,\rho_{- _{g}+ _{g}}\ ,
\end{equation}
where $\tan\alpha=\sqrt{2\eta c}/\left|\Omega_{c}\right|$. It is then shown to propagate as follows:
\begin{equation}
\left(\partial_{z}+\frac{\mathrm{i}\nu}{v_{g}\left(\alpha\right)}\right){\cal P}=\frac{4\mathrm{i}\eta}{3\Delta}\left({\cal P}-{\cal P}^{*}\right)\ ,
\end{equation}
The left hand side is the usual DSP propagation equation at a group velocity $v_{g}\left(\alpha\right)=c\,\cos^2\alpha$, and the right hand side factor leads to the FWM process parametric gain. Indeed, the DSP propagation is described by the same symplectic matrix as the probe field (see \ref{Appendix}).

\begin{figure}
\begin{centering}
\includegraphics[width=1\columnwidth]{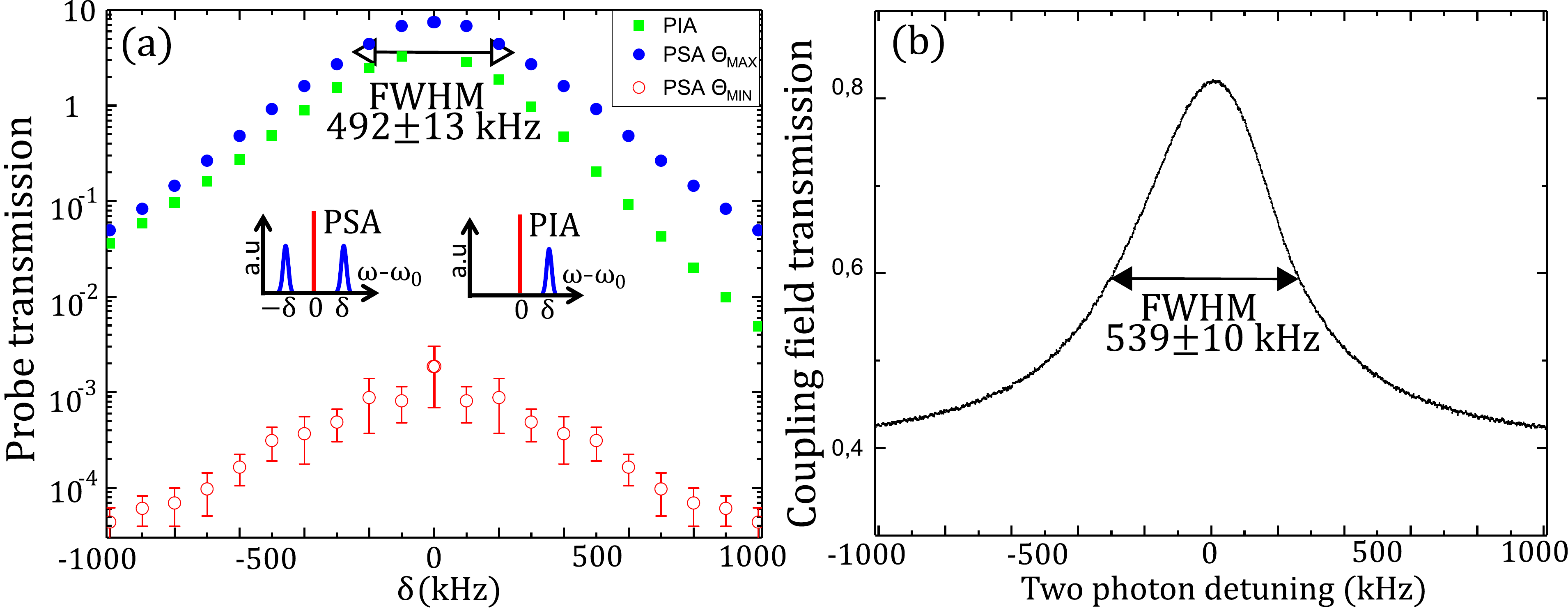}
\par\end{centering}
\caption{(a) Signal field transmission measured as a function of its detuning $\delta$ with respect to the coupling field frequency. Squares: PIA, without input idler field. Filled (empty) circles: maximum (minimum) PSA, with an input idler field. Error bars correspond to 1 standard deviation. (b) Measured CPT resonance for the coupling field. \label{Exp}}
\end{figure}

To experimentally test our interpretation of a CPT-enabled PSA process, we measured the probe minimum and maximum transmissions as functions of the detuning $\delta$ (see Fig.\,\ref{Exp}(a)). We compared the phase insensitive amplification (PIA) configuration, where the $+\delta$ signal field is sent alone, with the PSA configuration. In the former case, FWM spontaneously generate an idler field, leading to a phase insensitive gain of less than 3. In the latter case, PSA is observed with maximum gains up to 9 dB.

The coupling field CPT transmission bandwidth can be measured (Fig.\,\ref{Exp}(b)) by applying a tunable magnetic field along the propagation axis in the absence of a probe field: the ground levels are then Zeeman shifted, inducing a two-photon detuning, which cancels CPT. The full width at half maximum (FWHM) of the PSA and CPT profiles are comparable, stressing the fact that the coupling field power controls the bandwidth of both processes. One can notice that the CPT resonance does not reach full transparency at $\delta=0$: this is due to the large room-temperature $D_{2}$ transition linewidth $W$ leading to a residual absorption of $12\pm4\:\%$. These losses also explain that we experimentally have $G_{\mathrm{MAX}}G_{\mathrm{MIN}}<1$.

%In the situation where the signal field spectrum fits within the CPT bandwidth, one can derive a propagation equation for the dark state polariton (DSP) ${\cal P}$, which is a coherent superposition of the probe field $\Omega_{p}$ and the coherence $\rho_{- _{g}+ _{g}}$ between the dark and the bright states  \cite{Fleischhauer00}: 
%\begin{equation}
%{\cal P}=\cos\alpha\Omega_{p}-\sqrt{2\eta c}\,\mathrm{i}\sin\alpha\,\rho_{- _{g}+ _{g}}\ ,
%\end{equation}
%where $\alpha$ is a mixing angle defined by $\tan\alpha=\sqrt{2\eta c}/\left|\Omega_{c}\right|$. Using the Maxwell-Bloch formalism, one can show that this quasi-particle
%propagates as follows:
%\begin{equation}
%\left(\partial_{z}+\frac{\mathrm{i}\nu}{v_{g}\left(\alpha\right)}\right){\cal P}=\frac{4\mathrm{i}\eta}{3\Delta}\left({\cal P}-{\cal P}^{*}\right)\ ,
%\end{equation}
%where one recognizes on the left hand side the usual DSP propagation equation at group velocity $v_{g}\left(\alpha\right)=c\,\cos^2\alpha$. On the right hand side, one has a gain term that leads to the parametric gain generated by the FWM process in the system. One actually shows  \cite{SuppMat} that the DSP propagates with the same symplectic matrix as the signal field.

\section{Conclusion}

In this article, we have shown that the very efficient PSA process that we previously observed in a hot vapor of helium at room temperature \cite{Lugani16} is actually enabled by CPT. This process was demonstrated to provide a strong parametric gain as large as 9 dB for much lower optical depths ($\sim4.5$) than in usual alkali vapor setups \cite{Wang00,Lu98,Wang17}. Contrary to these previous works, we have a closed $\Lambda-$system which allows to fully exploit the nonlinearity enabled by CPT. Moreover, we derived a full analytical model to extract the transfer matrix of the probe. It well describes the experimental data and reveals an unusual scaling of the gain. Finally, an original physical picture of this effect could be derived using superpositions of atomic states.

The full transparency of the resonant $D_1$ transition allows for efficient FWM involving the detuned $D_2$ transition. Such a CPT-enabled PSA process should be associated to highly squeezed states generation, which will be addressed in a future work. Moreover, the propagation features of the DSP suggest the possibility to store and generate on-demand two-mode squeezed states of light, with the same atoms used recently to demonstrate coherent population oscillation based storage \cite{Neveu17}. Although this process is demonstrated in helium 4, our calculations and the advances on artificial atoms technologies make it possible to imagine systems designed to optimize it.
\ack
The authors acknowledge funding by Indo-French CEFIPRA agency, Labex PALM (Grant No. ANR-10-LABX- 0039-PALM), R\'egion IdF DIM Nano-K, and IUF.
\appendix
\section{Propagation equation of the signal}
\label{Appendix}
The analytical treatment developed in the main text (Eqs. (\ref{VN},\ref{eq04}) is
based on the Maxwell-Bloch equations.
The approximation framework is the following:
\begin{itemize}
\item the strongest nonlinear processes involving the $D_{2}$ transition
are isolated using a perturbative development to the first order in
${\cal O}\left(\frac{\zeta,\Gamma}{\Delta}\right)$.
\item we assume $\left|\Omega_p\right|\ll\left|\Omega_c\right|$, which legitimates a perturbative expansion at first order in $\Omega_p/\Omega_c$.
\item we assume $\nu\ll\zeta\ll\Gamma\ll\Delta$.
\end{itemize}
In that regime, using a formal computation software, we derive the
following propagation equation for the probe field (Eq. (12) of the main text):
\begin{equation}
\left(\partial_{z}+\mathrm{i}\frac{\nu}{c}\left\{ 1+\frac{2\eta c}{\left|\Omega_{c}\right|^{2}}\frac{1}{1-\mathrm{i}\nu/\zeta}\right\} \right)\Omega_{p}=\frac{4\mathrm{i}\eta}{3\Delta\Omega_{c}^{*}}\frac{2\Omega_{p}\Omega_{c}^{*}-\Omega_{p}^{*}\Omega_{c}}{\left(1-\mathrm{i}\frac{\nu}{\zeta}\right)^{2}}.\label{eq:PropaSignal}
\end{equation}
And using Eq. (\ref{eq:PropagStationnaireC}), it is possible to show that
\[
\Omega_{c}\left(z\right)=\Omega_{c}\left(0\right)\exp\left[\frac{4\mathrm{i}\eta}{3\Delta}z\right].
\]
In order to solve Eq. (\ref{eq:PropaSignal}), it is convenient to
get rid of this $z$-dependent phase shift of the coupling field by
introducing the new variable $\Omega_{p}^{'}\left(\nu,z\right)=\Omega_{p}\left(\nu,z\right)\exp\left[-\frac{4\mathrm{i}\eta}{3\Delta}z\right]$.
At first order in $\nu$, we obtain:

\begin{equation}
\left(\partial_{z}+\mathrm{i}\frac{\nu}{c}\left\{ 1+\frac{2\eta c}{\left|\Omega_{c}\right|^{2}}\right\} \right)\Omega_{p}^{'}=\frac{4\mathrm{i}\eta}{3\Delta}\left(\Omega_{p}^{'}-\Omega_{p}^{'*}\right).\label{PropaSignalPrime}
\end{equation}
The quantity $v_{g}\left(\alpha\right)=c\times\cos^{2}\alpha=c/\left(1+\frac{2c\eta}{\left|\Omega_{c}\right|^{2}}\right)$ is the usual group velocity of a light field in CPT (or EIT) conditions. Solving Eq. (\ref{PropaSignalPrime}) by considering the real and imaginary parts of $\Omega_{p}^{'}$ independently, one finally finds
\[
\Omega_{p}^{'}\left(L,\nu\right)=e^{-\frac{\mathrm{i}\nu L}{v_{g}}}\left[\left(1+\mathrm{i}\mu\right)\Omega_{p}^{'}\left(0,\nu\right)-\mathrm{i}\mu\Omega_{p}^{'*}\left(0,\nu\right)\right].
\] 
Then, going back to $\Omega_p$, one finally finds the transfer matrix in Eq. (\ref{eq:Matrix}), with the additional information that the propagation is at group velocity $v_g$.

\section{Propagation equation of the dark state polariton (DSP)}

Because EIT is occurring between the dark and bright states of the
system, one can then define the DSP by
\[
{\cal P}=\cos\alpha\Omega_{p}^{'}-\mathrm{i}\sqrt{2\eta c}\sin\alpha\rho_{+_{g}-_{g}}.
\]
Note that the above expression differs from the usual one \cite{Fleischhauer00} by the $+\mathrm{i}$ factor, which merely comes from the probe polarisation decomposition.

In the same approximation framework as above, the coherence between the dark and bright states writes 
\[
\rho_{+_{g}-_{g}}=-2\mathrm{i}\Omega_{p}^{'}/\left|\Omega_{c}\right|,
\]
so that ${\cal P}\times\cos\alpha=\Omega_{p}^{'}$. Using the latter relation to express Eq. (\ref{PropaSignalPrime}) in terms of $\cal P$ and $\cal P^*$, we get the following DSP propagation equation

\[
\left(\partial_{z}+\mathrm{i}\frac{\nu z}{v_{g}\left(\alpha\right)}\right){\cal P}=\frac{4\mathrm{i}\eta}{3\Delta}\left({\cal P}-{\cal P}^{*}\right).
\]

This equation and its complex conjugate can be solved so that we obtain $\cal P$ and $\cal P^*$  at $z=L$:
\begin{equation}
\left(\begin{array}{c}
{\cal P}\left(L,\nu\right)\\
{\cal P}^{*}\left(L,\nu\right)
\end{array}\right)=e^{-\mathrm{i}\frac{\nu L}{v_{g}\left(\alpha\right)}}\left(\begin{array}{cc}
\left(1+\mathrm{i}\mu\right)e^{\mathrm{i}\mu} & -\mathrm{i}\mu e^{\mathrm{i}\mu}\\
+\mathrm{i}\mu e^{-\mathrm{i}\mu} & \left(1-\mathrm{i}\mu\right)e^{-\mathrm{i}\mu}
\end{array}\right)\cdot\left(\begin{array}{c}
{\cal P}\left(0,\nu\right)\\
{\cal P}^{*}\left(0,\nu\right)
\end{array}\right).\label{eq:MatrixDSP}
\end{equation}
This equation coincides with Eq.\, (\ref{eq:Matrix}) up to an exponential factor due to the dispersive propagation of the DSP with the finite group velocity $v_g$.

\end{document}